\documentclass[aps,12pt,tightenlines,amsmath,amssymb]{revtex4}

\def\Eqref#1{Eq.~(\ref{#1})}
\def\littlespace{$\;$}
\newcommand{\be}{\begin{equation}}\newcommand{\ee}{\end{equation}}

\def\gtsim{\hbox{\kern.25em\raise.5ex\hbox{$>$}\kern-.75em\lower.5ex
   \hbox{$\sim$}\kern.25em}}

\def\taupass{\tau_\mathrm{passage}}

\begin{document}
%%%%%%%%%%%%%%%%%%%%%%%%%%%%%%%%%%%%%%%%%%%%%%%%%%%%%%%%%%%%%%%%%%%%%%
\title{Is ergodicity a reasonable hypothesis?}
\author{Bernard Gaveau}
\affiliation{Laboratoire analyse et physique math\'ematique, 14 avenue F\'elix Faure, 75015 Paris, France}
\author{Lawrence S. Schulman}
\affiliation{Physics Department, Clarkson University, Potsdam, New York 13699-5820, USA}
\email{schulman@clarkson.edu}
\date{\today}
%%%%%%%%%%%%%%%%%%%%%%%%%%%%%%%%%%%%%%%%%%%%%%%%%%%%%%%%%%%%%%%%%%%%%%
\begin{abstract} In the physics literature ``ergodicity'' is taken to mean that a system, including a macroscopic one, visits all microscopic states in a relatively short time. We show that this is an impossibility even if that time is billions of years. We also suggest that this feature does not contradict most physical considerations since those considerations deal with correlations of only a few particles.
\end{abstract}
%%%%%%%%%%%%%%%%%%%%%%%%%%%%%%%%%%%%%%%%%%%%%%%%%%%%%%%%%%%%%%%%%%%%%%

\maketitle

%%%%%%%%%%%%%%%%%%%%%%%%%%%%%%%%%%%%%%%%%%%%%%%%%%%%%%%%%%%%%%%%%%%%%%
\section{Introduction\label{s.intro}}
%%%%%%%%%%%%%%%%%%%%%%%%%%%%%%%%%%%%%%%%%%%%%%%%%%%%%%%%%%%%%%%%%%%%%%

Ergodicity has many faces. It started in physics, became mathematics, and remains important in physics as a supporting concept for the foundations of statistical mechanics. The original idea \cite{boltzmann1, boltzmann2} was that phase space averages should equal time averages \cite{note:precise}. The expectation is that as a system evolves in time it visits everywhere it can, so that if you average over time or if you average over possible locations in phase space, you get the same answer. The purpose of the present paper is to show that for macroscopic systems this idea is untenable.

We recall the general framework. For simplicity people generally assume that the system is in a box glued to the table \cite{note:astro}, meaning that the only constant of the motion is energy. Thus the energy surface, $E=H(p,q)$, defines where the system ``can'' go. As usual, $E$ is energy, $H$ the Hamiltonian, $p$ \textit{all} momenta ($N$ components if there are $N$ degrees of freedom) and $q$ all position coordinates.

In statistical mechanics one starts (conceptually, at least) with the microcanonical ensemble, namely the system is isolated and possesses a fixed total energy. The fundamental assumption---that \textit{all} states are equally likely to be occupied---is then considered a consequence of ergodicity. For example, another statement of ergodicity is that there is only one surviving constant of the motion, the energy, and again, this means the system goes everywhere consistent with that constraint.

Ergodicity has been questioned for physical systems. An example of a non-ergodic system is a collection of $N$ harmonic oscillators ($N>1$). The orbits fill a space of dimension $N$, while the energy surface is of dimension $2N-1$\@. There is then the famous KAM theorem (see for example \cite{poschel, zaslavskyBook, zaslavskyPhysicToday}) that says that for small non-linearity the structure is preserved, meaning that one should expect additional non-energy constants of the motion, for sufficiently small deviations from linearity (i.e., from harmonic oscillator dynamics). Does this mean that KAM tori ruin egodicity? Most physicists think not: in higher dimension there is what is known as Arnold diffusion that allows movement between the tori, and presumably visitation of \textit{all} regions of phase space.

But it doesn't work. What we show below is that the naive notion of ergodicity is absolutely, totally, with no doubt whatsoever, \textit{not} satisfied as an observed property of macroscopic physical systems. This has nothing to do with KAM tori. Our demonstration concerns an ideal gas. So there is no interaction and no chance of getting caught in special structures. What we do not understand is why this simple observation was not made a century ago.

For example in the authoritative text by Landau and Lifshitz \cite{landau} they say, ``\dots\ during a sufficiently long time the subsystem considered will be many times in every possible state.'' Although for them, ``sufficiently long time'' is not defined, it is clear that observational time intervals are considered and that this passage through ``every possible state'' will turn out to justify the foundations of statistical mechanics. We note incidentally two distinctions between our focus and that of the mathematical literature. First, demonstrations of ergodicity typically deal with a single particle, whereas we are concerned with the thermodynamic situation, with many particles. Second, the mathematician's ``$\mskip2mu t\to\infty$'' and the physicist's are two very different concepts. Thus, proofs of ergodicity---which are mostly limited to ``billiards,'' but are proofs nonetheless---also rely on the infinite time limit. As we show below, the time scales for exploration of ``all'' of phase space for macroscopic systems are vastly longer than the lifetime of the universe.

On the other side of the ledger, we will consider the practical implications. The normal behavior of gases does not seem to be affected by our arguments. In other words, if things are so terrible, why do the usual assumptions work?

Our main conclusion is that the assumption of ``equal a priori probability," postulating equal probability for all states that are in principle accessible, is far stronger than is supported by experience. This same conclusion has been drawn in connection with the ``special state'' theory of quantum mechanics \cite{timebook}, but in that case a more elaborate story is involved.

In the next section we show how far from reasonable the assumption of ergodicity is---even for non-interacting systems. Following that we consider some of the ordinary conclusions one draws about large systems, and find them to be justified.

%%%%%%%%%%%%%%%%%%%%%%%%%%%%%%%%%%%%%%%%%%%%%%%%%%%%%%%%%%%%%%%%%%%%%%
\section{Counting \label{s.counting}}
%%%%%%%%%%%%%%%%%%%%%%%%%%%%%%%%%%%%%%%%%%%%%%%%%%%%%%%%%%%%%%%%%%%%%%

We proceed in two steps. First we make ``reasonable'' arguments, then we make rigorous ones, or at least rigorous bounds.

Our question is, does the system visit all possible states? Take a mole of N$_2$ in a cubic meter at 300\littlespace K\@. At this temperature, the mean velocity (from $\frac32kT=\frac12mv^2$) is $v_\mathrm{mean}\simeq517\,$m/s (the N$_2$ mass is about $4.652\times10^{-26}\,$kg). The de Broglie wavelength ($\sim \hbar/M_{\rm{N}_2}v_\mathrm{mean}$) is about $4\times10^{-12}\,$m, so if we take ``being in the same state'' to mean that all molecules are within 10\littlespace nm of given locations, that would vastly underestimate the number of states (to say nothing of demands on nearly equal velocity). In a cubic meter there are $M=10^{24}$ boxes of size 10\littlespace nm. The number of ways to arrange these $N_A$ (Avogadro's number) identical molecules is the combinatorial coefficient $C^M_{N_A}\approx10^{2.9193\times10^{23}}$ (ignoring double occupancy, which would only make things worse). For the indicated mean velocity, the time for passing through one of these little boxes, $\taupass$, is about 1.93$\times10^{-11}\,$s. If the big bang took place 13.8\littlespace b years ago, then the number of states passed through since then is $2.25\times10^{28}$ or about %$10^{28.352}$\@.
$10^{28.4}$\@.
It follows that the fraction of states visited is
%$10^{28.352}/10^{2.9193\times10^{23}} = 10^{-2.9193\times10^{23}+28.352} =10^{-(2.9193-\epsilon)\times10^{23}}$
$10^{28.4}/10^{2.9193\times10^{23}} = 10^{-2.9193\times10^{23}+28.4} =10^{-(2.9193-\epsilon)\times10^{23}}$,
where $\epsilon$ would be on the order of 10 to the minus 20-something.

The conclusion that we draw is: visit all states? No way.

The above calculation can be criticized. We took liberties with the definition of state change and who says 10\littlespace nm defines a state? So we turn to quantum mechanics, specifically the quantum mechanics of an ideal gas. It is known that for moderate temperatures the entropy is~\cite{baierlein}
\be
S = N k \left[ \log  \left(\frac {V/N}{\lambda_\mathrm{thermal}^{3/2}}   \right)+5/2 \right]
\,,
\label{e.010}
\ee
with
\be
\lambda_\mathrm{thermal}  =   \frac{\hbar}{ \sqrt{M_{{\rm N}_2} k_B T/2\pi}}
\label{e.020}
\ee
the thermal wavelength. The entropy of the system described above (N$_2$, etc.) is about 181\littlespace J/K\@. The multiplicity of states is then given by ${\cal M}=\exp(S/k_B)\simeq 10^{1.31\times10^{25}}$\@. So it turns out that our first estimate got too small a multiplicity by two powers of 10 \textit{in the exponent of the exponent}.

How rapidly can the system explore this enormous phase space? There are a number estimates on the greatest rapidity with which a system can evolve into one that is orthogonal to itself. Most of the earlier bounds \cite{fleming} involved the energy spread, $\left\langle \left(H-\langle H \rangle\right)^2\right\rangle$, which meant we would have to make assumptions about the Hamiltonian. But more recently an estimate due to Margolus and Levitin \cite{margolus} uses only the total energy. The estimate is $\taupass=\pi\hbar/E$\@.

For the energy we simply use $\frac32 N k_B T$ which is about $3.74\times10^{3}\,$J\@. This really cuts down on the fastest possible transition time relative to our earlier estimates, and substituting, we get $\taupass\simeq 8.9\times 10^{-38}\,$s. But the prospects for ergodicity are nevertheless dismal. The number of states visited since the big bang is now $ 4.35\times10^{17}/8.9\times 10^{-38}\simeq10^{54.7}$\@. It follows that the fraction of states visited during this enormous time interval is $ 10^{54.7}/10^{1.31\times10^{25}}\simeq10^{-(1.31-\epsilon)\times10^{25}}$ with $\epsilon$ even smaller than it was in the earlier estimate.

%%%%%%%%%%%%%%%%%%%%%%%%%%%%%%%%%%%%%%%%%%%%%%%%%%%%%%%%%%%%%%%%%%%%%%
\section{Things that do work. \label{s.work}}
%%%%%%%%%%%%%%%%%%%%%%%%%%%%%%%%%%%%%%%%%%%%%%%%%%%%%%%%%%%%%%%%%%%%%%

Why then have the usual assumptions been so successful? This puts us on less secure ground than our basic demonstration, but we venture possible explanations.

\paragraph{A false lead.}
First we reject one idea that might seem a possibility. Go back to our first non-quantum, rough estimate. Suppose you changed the grain size. Is there a size such that the number of states visited actually equals the total available? The total volume is $V$, the number of particles $N$, the volume per particle is $\ell^3\equiv V/N$, the temperature $T$, and the observation time $t$\@. Take the new grain size to be $\lambda$\@. Then the number of grains will be $M=V/\lambda^3$\@. Now however since $\lambda$ may be large we need to deal with the possibility of multiple occupancy. The number of states will thus be the number of ways of partitioning $N$ objects in $M$ boxes, which is the number of ways of arranging $M-1$ dividers in a row of $N$ objects, so that the total multiplicity of states is ${\cal M}=C^{M+N-1}_N\simeq \exp\left((M+N)\phi(x)\right)$ with $\phi(x)\equiv -x\log x -(1-x)\log(1-x)$ and $x\equiv N/(N+M)$\@. The number of states visited is the observation time divided by the passage time, which we again take to be $\taupass=\lambda/v$ with $v$ the mean velocity at temperature-$T$\@. Setting the logarithm of these quantities equal implies
\be
(M+N)\phi(x)=\log \frac t\taupass=\log \frac {vt}\lambda
\,.
\label{e.030}
\ee
We go to a dimensionless variable $u\equiv \ell/\lambda$. Now recall that $x=N/(M+N)=1/(1+\ell^3/\lambda^3)=1/(1+u^3)$, so that \Eqref{e.030} can be rewritten as
\be
\frac{\phi(x)}x=\frac1N\log ( R u )
\,,
\label{e.040}
\ee
with $R\equiv vt/\ell$\@. As is evident, there are only two independent parameters, $N$ and $R$\@. A rough solution to this transcendental equation for the parameter range of interest ($N\sim N_A$) is $\lambda\sim \ell (N/w)^{1/3} $, where $w$ goes to zero, but in a complicated way. For $R>N^{1/3}$, $w\sim \frac{\log R}{\log N}-\frac13$, while for smaller $R$ it is roughly $w\sim 1/(3\log N)$\@. But the point is not the details: the point is that $\lambda$ grows to macroscopic sizes. For the conditions described earlier and for observation times on the order of $10^{-4}\,$s, one gets $\lambda\sim5\,$cm. This does \textit{not} characterize a microscopic state nor even one that is hydrodynamic (i.e., mesocopic).

\paragraph{An example that works.}
What we do believe is the reason that the problems of ergodicity have gone unnoticed is that almost all experimental and theoretical assertions concern low order correlations. We consider the simplest of these, the one-particle distribution function. Suppose a single particle is a ``billiard in a stadium'' \cite{note:billiard, bunimovich, dellago, benettin}. If the stadium is a 2-by-2 square capped at both ends by radius-1 half disks, the billiard is known to be ergodic with Lyapunov exponent $L\gtsim0.4$~\cite{note:nonergodic}. If the grains are roughly $(1/n)\times(1/n)$ in size (for some large number $n$) then a single particle will have occupied almost all the grains when $e^{L\tau}\sim 2n^2$ (the 2 is a rough correction for multiple occupancy). Thus $\tau$ is insensitive to grain size, growing only as $(2/L)\log n$\@. If one imagined a gas of such particles (interacting with the walls, not necessarily with each other), the stadium would develop near uniform density on the same time scale. Thus if our nitrogen molecules of the previous examples were in the right shaped stadium (and realistically one could also include their bouncing off one another as hard spheres) the overall density would rapidly become Poisson distributed. Once that is the case one recovers all the comfortable assertions, for example the fact that the pressure in a bicycle tire (measured, say, across a 1\littlespace mm sided square) would \textit{never} exhibit a 1\% fluctuation~\cite{note:never}.

\paragraph{Other signs of normalcy}
For model systems in which it is possible to solve the dynamics in the presence of severe constraints, the absence of a large fraction of available states may not affect the statistics. One instance where this has been demonstrated is a gas of free ``particles'' having cat map dynamics \cite{arnold, note:catmap}.  The unit square is partitioned into $n^2$, $(1/n)\times(1/n)$, grains and the relative grain occupancies, $r_k\equiv($number of particles in grain-$k)/($total number of particles$)$, are taken as the observables. The Lyapunov exponent for this dynamics is $L=\log \frac{ 3+\sqrt5}{2}\simeq0.96$\@. Taking as before the relaxation time to be $\tau=(1/L)\log(2n^2)$, for $n=10$ we get $\tau\approx5$, which is also found numerically. The constraint in \cite{timebook, effectwithwebandbriefbookreference} is that all points must lie in particular grains at particular times, say grain-$k_1$ at $t_1$ and grain-$k_2$ at a later time $t_2$\@. Thinking in terms of purely initial conditions, this allows only 1 out of $n^2$ of the possible states of the system to be realized, so there is nothing like ergodicity. Nevertheless, at times less than \hbox{$t_2-\tau$} there is no discernable difference in the variables $\{r_k\}$ from the unconstrained behavior. This has been illustrated in \cite{timebook} for the entropy ($-\sum r_k\log r_k$), but it holds for the $\{r_k\}$ as well.

Another example is stochastic dynamics for the Ising model. Here ergodicity is not a problem and is built into the system (the transition matrix is irreducible and any state is accessible from any other). However, we mention it to illustrate a related point. For a $100\!\times\!100$ lattice there are $2^{10\,000}\approx10^{3000}$ spin states, meaning that many conclusions drawn from stochastic simulations are based on small samples of the set of possible states. Often those conclusions are accurate, suggesting that random sampling (even with the inevitably flawed randomness of a computer) can be effective. But sometimes, for example for spin glasses or near criticality, such samples can be deceptive and special techniques are needed to make one's way through the state space. This suggests that a reason for our own successful use of the ergodicity hypothesis is not that \textit{all} states are entered, but rather that those that are visited are in some sense typical. Moreover (again thinking of the exceptions), even typicality may not be recognizable.

%%%%%%%%%%%%%%%%%%%%%%%%%%%%%%%%%%%%%%%%%%%%%%%%%%%%%%%%%%%%%%%%%%%%%%
\section*{Acknowledgements}
%%%%%%%%%%%%%%%%%%%%%%%%%%%%%%%%%%%%%%%%%%%%%%%%%%%%%%%%%%%%%%%%%%%%%%
We are grateful to the Max Planck Institute for the Physics of Complex Systems, Dresden, for its kind hospitality.

%%%%%%%%%%%%%%%%%%%%%%%%%%%%%%%%%%%%%%%%%%%%%%%%%%%%%%%%%%%%%%%%%%%%%%

\begin{thebibliography}{21}
\bibitem[{not({\natexlab{a}})}]{note:precise}
\bibinfo{note}{There are volumes devoted to precise definitions of ergodicity
  and related concepts. Our approach will be heuristic.}
\bibitem[{not({\natexlab{b}})}]{note:astro}
\bibinfo{note}{If there's good reason for allowing another constant of the
  motion, like angular momentum (for astrophysical applications), that too is
  part of the specification of where the system ``can'' go.}
\bibitem[{not({\natexlab{c}})}]{note:billiard}
\bibinfo{note}{A ``billiard in a stadium'' is a point particle moving in two
  dimensions, confined by walls, but otherwise free. When reaching a wall it is
  specularly reflected.}
\bibitem[{not({\natexlab{d}})}]{note:nonergodic}
\bibinfo{note}{Even if one does not have ergodicity, just bouncing off hard
  walls will access all grains for almost all initial conditions, but
  assertions and estimates here would get us too deeply into number theory.}
\bibitem[{not({\natexlab{e}})}]{note:never}
\bibinfo{note}{The meaning of ``never'' is easy to calculate. The relevant
  volume is $10^{-3}\!\times\!10^{-3}\!\times vt_\mathrm{observe}\,$m$^3$\@.
  Taking the observation time, $t_\mathrm{observe}$, to be $10^{-2}\,$s and
  (under these conditions) the number of particles per m$^3$ to be
  $2.6\times10^{25}$, then there are roughly $10^{20}$ molecules in the volume.
  The relative standard deviation is then about $10^{-10}$ so that a 1\%
  fluctuation represents $10^{8}$ standard deviations, leading to a probability
  of about $10^{-28}$\@. This means that if you took one of these measurements
  every second since the big bang, you would have about 1 chance in a thousand
  of finding a 1\% error\dots\ once.}
\bibitem[{not({\natexlab{f}})}]{note:catmap}
\bibinfo{note}{The cat map sends the unit square into itself. The point $(x,y)$
  maps into $(x+y,x+2y)$ both sums taken modulo 1\@. The statistical mechanics
  model in the text is a ``gas'' of many such particles.}
\end{thebibliography}
\end{document}